Theoretical Foundation for Research in Communication using Information and Communication Technology Devices in Healthcare: An Interdisciplinary Scoping Review

*Arun Keepanasseril, BDS, MDS, MSc
McMaster University
1280 Main Street West
CRL Building
Hamilton, Ontario
Canada L8S 4K1
(647) 787-7352
keeppaa@mcmaster.ca

Kathleen Ann McKibbon, BSc, MLS, PhD
Associate Professor
Department of Clinical Epidemiology & Biostatistics
Director, eHealth MSc Program
McMaster University
1280 Main Street West
CRL-132
Hamilton, Ontario, Canada L8S 4K1
(905) 525-9140 x 22803
mckib@mcmaster.ca

Alfonso Iorio, MD, PhD, FRCP(C)
Associate Professor, Department of Clinical Epidemiology and Biostatistics
McMaster University
1280 Main Street West
CRL-140
Hamilton, Ontario, Canada  L8S 4K1
(905) 525-9140 x 22421
iorioa@mcmaster.ca

* Corresponding author

### Abstract

**Introduction and Background**

In spite of substantial investments in technology and training of personnel, healthcare is yet to improve its safety records substantially. Faulty communication between team members is one of the most important factors preventing substantial improvement in patient safety. Industries such as aviation, nuclear power and defense have been able to improve their safety record by adopting theory and model based solutions. In contrast, healthcare's thrust towards modern communication devices is largely devoid of theoretical foundation. The objective of this scoping review is to compile communication theories, frameworks, and models used by high risk organizations outside healthcare to study and resolve workplace communication issues.

**Methods**

The healthcare databases searched included Medline, CINAHL, EMBASE, and PsycInfo. In addition, we searched engineering and science literature to include articles in the fields of information sciences, computer sciences, nuclear power generation, aviation, the military and other domains such as sociology that address the science and theory of communication. Comprehensive searching was also done in the communication studies literature. We also reviewed conference proceedings and grey literature and conducted citation tracking.

**Results**

Our initial systematic search yielded 15,365 articles. Hand searching and reviewing references resulted in a set of 181 articles. One hundred and forty four full text articles were read and 40 of them were selected to be included in the review. We were able to identify 14 theories and 12 models which could be applied in hospital communication research.

**Conclusions**

Our review was able to identify a number of theories and models that may be applicable in furthering hospital communication research. However, it must be noted that most of them have not yet been applied in biomedical research in hospital communication and as such their applicability can only be suggested--a gap which future research may be able to address. Formulation of a custom model representing the unique features and complexities of communication within hospitals is recommended.

**Summary Table**

**What was already known on the subject:**

- Improvement in patient safety is not satisfactory despite substantial recent efforts.
- Faulty communication is a major cause for adverse events in hospitals.
- Information critical organizations such as aviation, nuclear power and defense have improved their safety records by studying their problems using a foundation based on theories and models.

**What this study adds to this knowledge:**

- Compiles a list of theories and models that may be applicable to hospital communication research and development.
- Suggests formulation of a new model customized for hospital communication as none of the models have been formulated with the unique needs of professional communication within healthcare teams.

**Introduction and Rationale**

Despite substantial investments in technology and training of personnel particularly in the past decade "we are no further ahead in improving patient safety and hospital complications are no less frequent now than in the 1990s"[1], [2]. In fact, some reports suggest that hospitals have become more dangerous[1], [3] . Much of the blame for this frustratingly slow progress in improving patient safety may be attributed to faulty communication within and between healthcare teams.

Faulty communication between the members of the care team has been identified as one of the major factors contributing to adverse events (AEs) in hospitals [3]-[7]. Various studies have linked as much as 25% to 81% of AEs in hospitals to errors in communication [8]-[10][Kesten]. Communication errors are consistently the most frequent contributor to over two - thirds of sentinel events reported to the United States Joint Commission [JCAHO]. Nearly half of the AEs reported to the health and disability commission in New Zealand were attributed to communication errors [7], many of which are likely communication problems. In a review of 14 000 hospital admission in Australia, communication issues were identified as the most common cause of clinical errors [11]. In addition to patient harm, communication errors in hospitals are a huge drain on economic resources – according to one estimate medical errors cost North American hospitals approximately 20 billions USD in 2008 in direct costs alone [12].

Introduction of state-of-the-art communication devices is a major component of healthcare's efforts to improve interprofessional communication. Hospitals in United States, Europe, Canada and elsewhere are heavily investing in smart phones, tablets and similar devices to enhance interprofessional communication[13] [need citation]. However, improvements in care have not been very much encouraging: a recent systematic review on use of smartphones in hospitals concluded that "there is limited evidence demonstrating that the use of technology leads to direct improvements in either clinician efficiency or patient care" [14]. Evidently, substantial scope for improvement exists in this area.

Healthcare is only one of many domains where operational errors carry high risk of harm to human life. Communication errors in information - critical High Risk organizations (HROs) such as aviation, nuclear power and defense can be fatal for the personnel involved. Each of those domains is safer today because they studied their communication challenges on a firm scientific and theoretical foundation and implemented customized solutions unique to the needs of their situations and settings. In contrast with aviation, nuclear power and defense, healthcare's thrust towards using modern communication devices is largely devoid of a theoretical foundation[15].

We propose that healthcare needs to study its communication challenges and the effects of each new device or change in work practices on communication effectiveness and safety. This study must be based on theories and models so that we can emulate the success of other HROs mentioned previously. Our initial literature search could not locate any peer-reviewed scientific reviews covering our focus area of the theory, models or frameworks of health professional communication using electronic devices. This scoping review compiles communication theories, frameworks, and models used by HROs outside healthcare to study and resolve workplace communication issues. The theories and models identified in this review will cover a broad range of basic and applied sciences such as psychology, human factors engineering, organizational behavior, neuro-ergonomics, and cognitive psychology among others. To our knowledge, such a review has not been performed.

**Methods**

Few studies of hospital communication using new electronic devices are built on theory, models, or frameworks regardless of the domain. This lack is consistent with other aspects of e-Health as the domain is new and lacks a theoretical basis [15]. Therefore we conducted this review using scoping review methods. Scoping reviews are most effective when a highly specific body of literature does not exist [Danielle, 2010]. Scoping reviews rely on rigor to identify, collect, extract and analyze data as do other systematic reviews. However, the literature in scoping reviews is often sparse, broadly based, and comprised of studies of varying research methods. Scoping reviews identify strengths and gaps in the evidential base and summarize them setting the stage for further work.

The scoping review framework originally proposed by Arksey and O'Malley[16] and later modified by Levac, Colquhoun and O'Brien[17] was used in this review. Wherever the unique nature of the project warranted a modified approach to the prescribed methods, they were fine-tuned accordingly and described in this document.

Because information on the theories and models is often included in the body of articles, standard searching methods based on indexing terms and notation in titles and abstracts were not likely going to be effective for article identification. Therefore we conducted a substantial amount of hand searching and bibliography checking. Screening was also complex and time consuming and required review of many full text articles in print and electronic format.

Our project has 4 basic concepts: location of the communication (e.g., hospitals or nuclear power plants); theories, models or frameworks related to communication; the communication itself; and the use of new information communication technologies (ICTs).To be included in this review the article must discuss one or more theories, models and frameworks. These must have been applied to communication systems, used to propose either a new or modified communication system or if the authors have proposed or used a theory, model or framework to evaluate communication issues in any domain of interest (health care, nuclear power, military, or aviation).

Our searching sought to capture 3 groups of studies or articles. The first category describes communication theories, models and frameworks that have been applied to interprofessional communication in any of our settings of interest. These articles are more theoretical and often do not have study data. The second group consists of articles that contain descriptions of methods of communication error analysis in our settings of interest using the new ICTs and use or mention a theory, model or framework in the article. From our initial review of the literature we found that the theories, models and frameworks we sought were often embedded within individual studies of communication. The third group consists of stand-alone articles (not experimental studies) that describe these theories, models and frameworks. Hypothetical, abstract and purely academic theories, models, or frameworks as well as those that are judges to be solely pertaining to interpersonnel communication without the use of ICTs were not included.

The healthcare databases searched included Medline, CINAHL, EMBASE, and PsycInfo. In addition, we searched engineering and science literature to include articles in the fields of information sciences, computer sciences, nuclear power generation, aviation, the military and other domains such as sociology that address the science and theory of communication. Comprehensive searching was also done in the communication studies literature. We also reviewed conference proceedings and grey literature. We conducted citation tracking and bibliography checks. Our searching was iterative in nature and each database and domain needed tailored searching strategies. Our search strategies are available from author on request.

Titles and abstracts were reviewed in the first round of screening and full texts in the second round. Screening and data extraction were done in duplicate.

Data extraction was done on theories, models and frameworks of interest; the communication processes studied; professionals involved in the communication; study type; ICT technologies used; publication year; outcomes; name and any synonyms of the theory, model or framework; its relation to other theories (e.g., if the new theory is developed from a previous theory); subject domain; attributes and how it was applied to communication using new ICTs.

### Table 1: List of Databases Systematically Searched

| Ovid MEDLINE (1948-Oct 2011) | EMBASE | Psycarticles (1806-Oct 2011) |
| OVID Health and Psychosocial Instruments | IEEE | Engineering Village |
| CINAHL | Business Source | Military and Intelligence |

|  | Complete | database |
|---|---|---|

**Table 2: List of Databases and Sources Hand Searched**

| Scholar's Portal | Business Source Complete | International Journal of Aviation Psychology |
|---|---|---|
| Google Scholar | Proquest | International Journal of Human-Computer Interaction |
| Tayler and Francis | Science Direct | OVID Health and Psychosocial Instruments |
| Cognitive Systems Research | Leadership in Health Services | ACM Special interest group - HCI |
| Computer Supported Cooperative Work | Cognition, Technology & Work | Journal of Loss Prevention |
| Computers in Human Behavior | Social Science & Medicine | Human Resource Management Review |
| International Emergency Nursing | Nursing Times Research | Journal of Communication Management |
| International Journal of Nursing Studies | Internet Research | Campus-Wide Information Systems |
| Journal of Managerial Psychology | Wilson Web | Corporate Communications: An International Journal |
| Journal of Clinical Psychology in Medical Settings. | Multicultural Education & Technology Journal | International Journal of Operations & Production Management |
|  |  | Business Process Management Journal |

**Results**

Table 1 lists the databases which were systematically searched. Table 2 lists the databases and literature sources electronically or manually hand searched.

Our initial systematic search yielded 15 365 articles. Hand searching and searching back references resulted in a set of 181 articles. One hundred and forty four full text articles were read and 40 of them were selected to be included in the review. We were able to identify 14 theories and 12 models which could be applied in hospital communication research. Primarily for ease of comprehension and clarity of presentation, a system of categorization of the theories and models located by this review has been proposed. The theories have been classified in to three groups based on their broad area of focus – interpersonnel, organizational and system-wide. Models have been classified in to two groups based on their content in to either mechanical / mathematical or psychosocial / composite models. They are listed in Tables 3 and 4 respectively. Figure 1 illustrates the search and selection process. The discussion section briefly describes each theory and model and then suggests in what situations the model would be useful in hospital communication studies using ICTs.

**Figure 1 : Diagrammatic Representation of the Search and Selection Process**

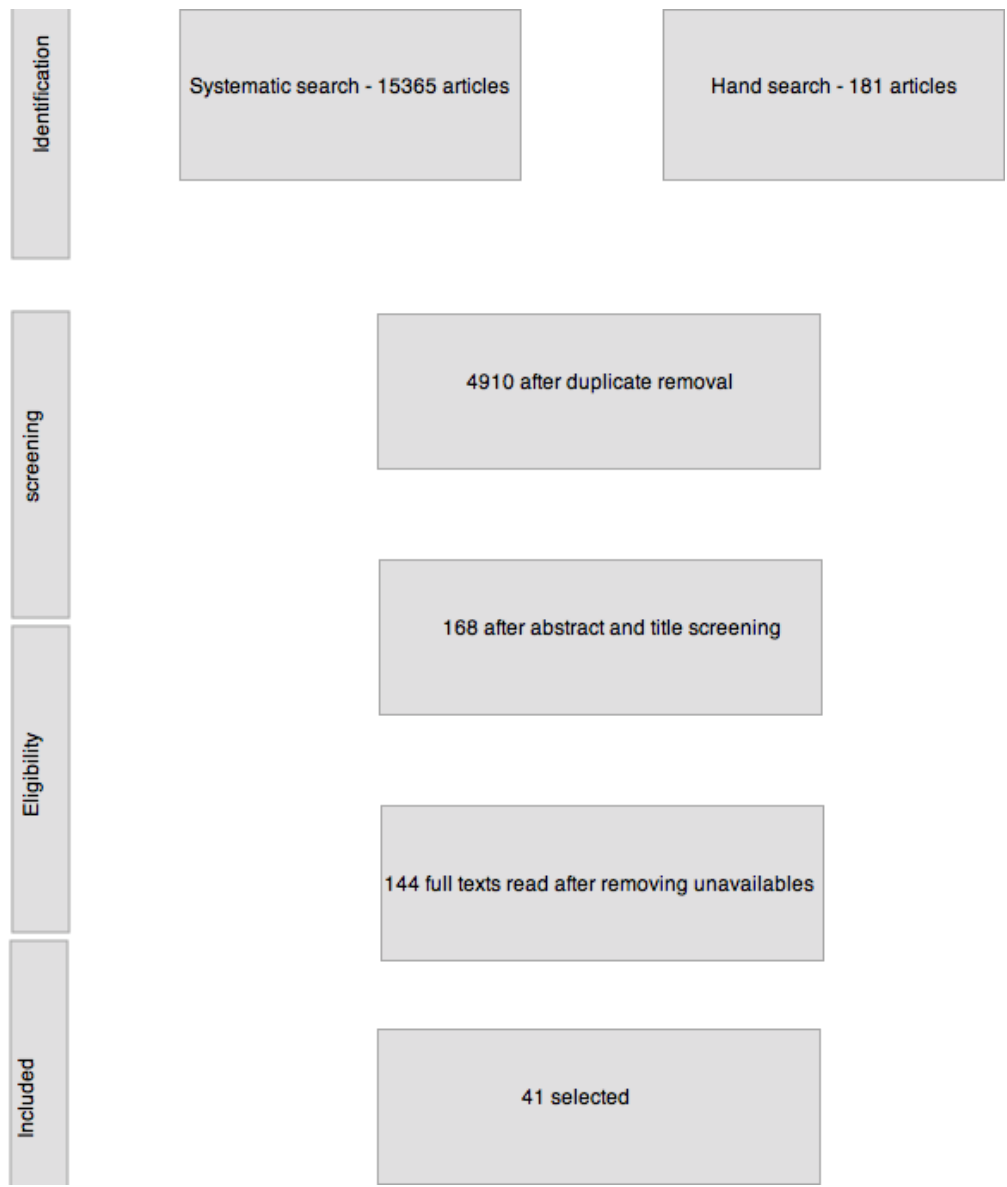

**Table 3: Classification of the theories according to their focus**

| System level theories | Organizational level theories | Interpersonnel level theories | Other |
|---|---|---|---|
| Information theory | Taylor's theory | Uncertainty reduction theory | Wicked problems theory |
| Network theory | Chaos theory | Humanistic communication theory | Task artifact cycle |
| Social network theory | Risk communication theory | Communication accommodation theory | |
| Likert's four systems theory | | | |

**Table 4: Classification of models according to their content**

| Psychosocial and / composite models | Mechanical . mathematical models |
|---|---|
| Systemic interaction model | Shannon weaver's model |
| Schein's cultural model | Action interaction and transactional model |
| Westrum's model | Helical model |
| SMCR model | Schramm's model |
| Andersch Staats and Bostom model | Circular model |
| Multiple resource model | |

**Discussion**

Healthcare has traditionally been reluctant in adapting to changes and was slow in embracing digitization and automation of work processes. However, that has changed in the past decade and a half - today healthcare is passing through a plethora of changes in relation to technology and operational aspects. One notable feature of these unprecedented changes is emphasis on improving teamwork and enhancing patient safety and quality of care. Organizational culture in large and small hospitals alike is also changing from a culture of individual culpability to a systems approach based firmly on teamwork. A notable feature of this change to teams is that ICT remains at the center of all these changes. Mere implementation of state of the art devices can only provide marginal or temporary gains at best.

Before we can automate a task, we must first model it. As with other areas of eHealth, the fast paced ICT adoption in hospitals is largely lacking in a sound theoretical basis. Healthcare needs to study its unique processes and challenges and adapt existing devices to it, or even develop custom solutions. Experience of other safety critical industries, especially those of aviation and nuclear power generation has demonstrated that applied research with strong theoretical underpinnings is mandatory if significant gains in safety are to be achieved. The prime objective of this review was to create a repository of models and theories that can serve as a starting point for future work in the field of hospital communication. A theoretical model can enhance understanding of complex situations. In the absence of such models interventions might end up addressing inappropriate variables or tackle only a small portion of the variables in question. Current approaches to improving communication within hospitals are rarely theory driven and are mostly based on implementation of state of the art communication devices. Undoubtedly, these devices have enhanced usability and availability of communication channels but generally they fall short in overall improvement of patient safety and teamwork. A theory will provide the basis for judging whether all the fundamentals are in place when studying hospital communication or implementing new technologies.

A brief discussion of the theories and models identified by this review is provided. For the sake of presentation the theories have been grouped under three headings based on their primary focus or level of the intent of the model: systems, organizations, individuals or others. The models are also divided in to three groups based on their core content: psychosocial, composite, or mechanical and mathematical models With each description, salient features of the model that apply directly to hospital communication with new ICTs are highlighted.

### *Theories*

### System level theories

Systems theory is the term used to denote a group of theoretical concepts originated from systems thinking which has a wide variety of interpretations and adaptations in various fields. In systems theory a system is considered to be a set of objects or entities that interrelate with one another to form a whole. Systems can be closed (one that has no interchange with its environment) or open (one that accepts matter and energy from, and send it to its environment). Systems consist of objects, their attributes, internal relationships and environment. Systems tend to be embedded within one another (hierarchical). Some other key properties of a system are cybernetics and homeostasis or balance. Cybernetics deals with the ways systems, along with their subsystems, modulate their affect and make necessary adjustments[4]. Many of the departments in a large hospital are so diverse in the scope of their objectives, operative procedures and team composition that each of them may be considered semi-autonomous units, as for example, an emergency room and the operating room complex in a large hospital. Ever increasing levels of specialization sometimes ensures that those units are often studied in relative isolation. From the systems theory perspective every unit constitutes a set of objects that interrelate with one another to form a whole, the hospital itself. Admittedly each unit has its own set of attributes, but patient safety research addressing communication issues may be enhanced by careful study of internal relationships of the various 'subsystems' inside the larger hospital system.

**Information Theory**

Information theory focuses on measurement of information. This theory deals with the quantitative study of information in messages and flow of information between senders and receivers. According to this theory, information is the level of entropy, the unpredictableness in a system. Higher levels of entropy are associated with more information, and vice versa (negantropy). Information can also be viewed as the number of messages needed to completely reduce the uncertainty in a situation. The higher the complexity of the situation, the higher will be the number of alternatives. Another concept in information theory is redundancy. Redundancy is a measure of the predictability of information. A certain amount of redundancy is necessary for complete understanding. Redundancy also enables systems so that they can be self-checking and with even more redundancy to become self-correcting[4]. According to information theory, some amount of redundancy is good, but the limit is not known. As the hospital team members coordinate progressively and form a shared mental model of the patient the amount of entropy is reduced. It follows that messages between hospital care teams must contain the right amount of information with some redundancy to convey the intended message. Incomplete or highly redundant messages can result in patient harm.

**Network Theory**

Essentially a network is a representation of how information is created and flows through an organization. According to network theory the members of an organization communicate and incorporate the results of the communications to produce some end product by using energy, information and materials from the environment. Two types of information can be present in an organization - absolute and distributed. Absolute information deals with what is known while distributed information deal with who knows it. It is important to know who knows what because the mere presence of information does not guarantee that it is available to all who need to know the information and may result in either information not reaching the intended personnel or rendered redundant by the time it reaches them.

One of the prime objectives of using modern communication devices for communication inside a hospital is to dissipate distributed information. Absolute information can be found embedded in the system as procedure protocols, automated alerts or others. However, distributed information is more loosely held by the personnel as this information needs to be delivered to whoever needs such information. Network theory can be used to evaluate communication dynamics within each subsystem.

**Social Network Theory**

Social network theory postulates that "technical work processes as well as social systems need to be considered and optimized to improve organizational performance". A social system comprises elements such as the employees and their practices, mental constructs and interactions. The tools, devices, materials and techniques by which work is performed make up the technical system. The social network theory holds that a balance is needed between technical and social systems for good communication.

Context influences the behavior of individuals in an organization. According to socio technical theory, technology is an important determinant of organizational context. Changes in technical systems will create changes in the social system. Many such changes may solve several existing problems but at the same time may bring about new kinds of errors. It is essential to study how each subsystem is influenced by context and the effect of those changes on other subsystems[18].

Sociotechnical theory can be easily applied to hospitals. Social system inside a hospital consists of doctors, nurses, pharmacists, other professionals and employees with their own behaviors, activities, skills, attitude, organizational and professional culture and their interactions. Communication devices such as pagers, cell phones and medical records form the artifacts in the system.

The sociotechnical theory implies that before introducing new devices and practices it is important to study their impact on the system as a whole. This is a very important point because many hospitals are in the process of upgrading their communication systems by introducing new devices such as Blackberrys, iPads and iPhones. The usage of these devices can affect changes in the sociotechnical system of the hospital which in turn will affect the hospital's safety performance. These new devices change the way team members communicate with each other. Therefore the existing communication practices need to be fine-tuned to accommodate the changes.

**Likert's Four System Theory**

Likert's subsystem theory proposes 4 subsystems to characterize an organization: exploitative authoritative, benevolent authoritative consultative and the participation management system. Communication functions as the intervening variable. Little communication happens in an authoritative system; communication channels are used only to inform the workers of a decision. In other words, all communication is 'upline'. Likert believed that the participatory management system was the best among the four as it encourages the formation of cohesive teams. Both 'upline' and 'downline' communication are present in the participatory management system. Effective and appropriate communication in an organization elicits cooperation and higher productivity on the part of the employees[19].

**Organisational level Theories**

Modern hospitals are large sized organisations and therefore it is very important that theories that deal with communication as an organisational function be considered.

**Theory of Organizational Communication**

In Taylor's Theory of Organizational Communication text is the content of an interaction and conversation is the interaction between two or more individuals. To understand the complexity of interaction between text and conversation, text needs to be translated into conversation and conversation back into text. Mere translation of text and conversation to each other does not guarantee that the intent and context are conveyed during communication. Often the same text may be translated by the sender and receiver in contrasting ways creating a gap between the intentions of the people involved in a conversation. This gap, dispensation, is explained in terms of degrees of separation. The separation can range from first degree as in a face to face conversation wherein the speaker's intention is accurately translated and embedded in the conversation to sixth degree when text and conversation are standardized and diffused through media for general consumption as in a television news item or a public leaflet.  According to Taylor, personal interactions inside an organization shape an organization. Concepts of text, conversation, intent, context, dispensation and degree of separation provide a framework to analyze communication[19].

Communications in a hospital can be viewed in terms of degree of separation. Face to face consultation between doctors and nurses is an example of first degree separation. It is the best mode of communication because efficient translation of text (content) and conversational interaction occurs??. In hospitals it is a common practice for the senior member of the team, usually the consultant, to be briefed by one of the team members. Second degree of separation occurs in such cases of communication. In the case of transcription of a physician's case notes by a transcription professional, separation is of the third degree. Professional communication between physicians is an instance of separation of fourth degree. Introduction of ICT-enabled devices change the communication patterns and organizational culture in a hospital[19].

## Chaos Theory

Chaos theory is not a theory of communication in the strictest sense but it represents "broad conceptualization of both organization and crisis that moves beyond the traditional crisis communication and public relations frameworks" [20]. "Sensitive dependence on initial conditions" also known as the butterfly effect is the most basic feature of chaos[20],. Stewart argued that chaos "produces a tiny change in the state of the atmosphere[20]". This small change in the atmosphere can produce catastrophic results over a period of time[2], [20]. Naturally occurring self-repeating patterns in a complex system are termed as 'fractals'. Those features or components of a system which maintain order are called 'attractors'. Minor failures in communication processes such as failures to transmit or receive warnings, improper interpretation and misreporting of messages can trigger major crises. This triggering aspect of Chaos Theory is applicable to hospitals also. Failure to report or acknowledge critical information can endanger patients' lives. Chaos Theory also serves as a means to explain the complexity of healthcare services [3], [21], [22].

## Risk Communication Theory

Risk communication theory is a subset of communication theory which aids in understanding how risks and crises can change the customary rules of communication [1], [3], [5]-[7]. This theory addresses the problems raised in the "exchange of information about the nature, magnitude, significance, control, and management of risks" [1], [9], [10]. It also deals with determining the strength of channels through which information about the risk is transmitted. Little agreement exists regarding the best method to transmit communications regarding risk. Basically the method is chosen based on the recipient. For example, emails, paging and text messages are some of the strategies used by hospitals for informing healthcare professionals about a potential nosocomial infection. Lack of information regarding context is one potential pitfalls of electronic transmission. According to Baruch Fischhoff, no single best technology is available to employ in risk communication; all technologies have their own shortcomings and advantages. Each of the technologies also has negative consequences for some segments of population[23] It is advisable to weigh the benefits and risks for a given situation before choosing one communication method over another.

## Interpersonnel level Theories

Even though hospital communication is predominantly professional and impersonal at the most basic level it remains an exchange of information between individuals at various levels. Therefore we have included certain theories that may be relevant from that perspective. Theories that deal exclusively with communication of a personal nature were not selected in this review.

## Uncertainity Reduction Theory

In Uncertainity Reduction Theory, Berger and Calabrese attempted to model the processes through which communication is used to reduce uncertainty in initial interactions between strangers. Although traditionally uncertainty reduction theory has been mostly used to analyze uncertainty in continuing relationships and in intercultural relationships this theory has been applied in the area of organizational communication also. The most important observation in this regard is made by Michael Krammer who stated that the Uncertainty Reduction Theory has implications for exploring communication as a means of resolving incompatibilities among cognitive structures, experiences and behaviors in various settings [4], [7]. Communication between members of healthcare teams can be regarded in terms of the level of uncertainty regarding a particular piece of information which is needed to care for the patient. The higher the level of cognitive uncertainty, the greater will be the amount of communication required for care coordination. The presence of team members who are new and unfamiliar with the patient's protocol and details can increase the amount of cognitive uncertainty and therefore, the amount of communications required.

**Humanistic Communication Theory**

Humanistic communication theory cautions against dehumanizing communication which occurs when the communication process is stripped of unique human characterization. Dehumanization occurs when one speaks to people as if they were "things" rather than human beings resulting in their devaluation[8].In the context of communication in hospitals, electronic communication devoid of real contact between providers is prone to 'dehumanizing' and can result in loss of context and situational awareness [8], [11].

**Communication Accommodation Theory**

Convergence, divergence and maintenance are three different concepts in the Communication accommodation theory. Convergence occurs when individuals adapt to each other's speech by means of a wide range of linguistic features whereas divergence occurs when people try to accentuate communicative differences. The third concept, maintenance, occurs when an individual's communicative patterns remain stable throughout the interaction. Convergence can be either upward (i.e., towards a socially accepted form of communication or speech) or downward (i.e., away from a socially sanctioned form of communication). Convergence can also be full, (i.e., the individual actually matches the communicative behavior of the other), partial or even hyper or crossover (i.e., the individual goes beyond the behavior of another on a part of or a particular dimension). Convergence can be unimodel (for example, converging on the dimension of vocabulary but not on the dimensions of accent or speech rate), multimodal (for example, converging on several dimensions of communicative behavior), symmetrical where both parties and the interaction attempt to converge towards each other or asymmetric where only one party in the interaction attempts to converge. These concepts need to be considered when new communication devices are given to healthcare team members. The mere provision of these devices may not improve the teamwork[4].

**Other Theories**

Wicked Problem Theory and Task Artifact Cycle are not theories of communication in the strictest sense but they are included in this review because their application can provide a unique perspective in studying the problem of hospital communication using new technologies.

**Wicked Problems Theory**

Wicked problems are those that are "ill-defined, ambiguous and associated with strong moral, political and professional issues" [18]. These problems are characterized by their dynamic nature. They exist in a complex interactive environment and are difficult to solve. Typically it is also difficult to know when a solution has been reached in these wicked problems [18].

Therefore, one may view hospital communication as a wicked problem. Wicked problems are better dealt with from socio-technical angle because this provides a complex framework in which to accommodate the multiple facets of the problem. It may be argued that the problems of hospital communication do not possess all the features of a typical wicked problem. But a case can be built up for its consideration as wicked if the sociotechnical angle of the problem is considered.

**Task Artifact Cycle**

'Task-artifact cycle', is a process by which technological artifacts are designed to support existing tasks[24] These tasks in turn are modified by the new technology, resulting in the need for redesigned artifacts. Multiple cycles of redesign can happen in communication [25]. The crux of this theory is that tasks and artifacts co-evolve. For example, to accomplish a task, a set of artifacts may need to be designed. These artifacts may in turn change the way the tasks are performed in that the artifact opens up new possibilities or constraints. Thus the artifact may end up changing the task. This series of changes does not stop there. The process can continue in a cyclical fashion endlessly. In simpler terms, introduction of a tool to facilitate a task will change the way the task is performed. This in turn will create new needs and necessitate changes in the tool. In terms of communication, this cycling can mean that communication process improvements and device designs will continue in a seemingly endless spiral of versioning in which processes and devices influence each other and modify mutually. This happens because requirements never stabilize and the introduction of the device into the process changes the process itself which in turn necessitates device refinement. As easy as this principle is to understand from a social science point of view, it poses considerable challenges to information systems design. That is because ICT can respond to finite predefined situations only. This places information systems in a position of disadvantage when it comes to keeping up with the ever changing human requirements. The rule based behavior of systems is not sufficient to deal with human and evolving situations [26]. According to the task artifact theory, ICT can even change the way patient care is delivered. For example, after electronic medical record system implementation physicians and nurses find that a large portion of their time otherwise spent in direct patient contact is taken up by record keeping. It is increasingly common to hear a physician complaining about this unexpected demand. One way to reduce the time needed for updating electronic records is by improving its usability. Usability requirements will then change the electronic medical record software presentation. This series of changes, according to the task artifact theory, is bound to be a virtually never ending spiral.

### Models

According to Mortenson, "In the broadest sense, a model is a systematic representation of an object or event in idealized and abstract form. Modelling a problem may not provide answers readily but allows us to ask questions"[27]. Twelve models were identified in this review (Table 2). For the sake of presentation they were divided in to three groups based on their core content and are describedin the following section. With each description, salient features of the model that apply directly to hospital communication with new ICTs are highlighted.

**Psychosocial and / Composite Models**

Unlike the earlier mechanical models and perhaps due to the increase in complexity of communication process several communication models developed in the latter half of 20[th] century attempted to portray communication as a composite process with contributions from a wide variety of domains such as psychology and organizational behaviour as well as fundamental concepts from historical mechanical models. Some of these models are described below.

Psychological models focus on personal characteristics of individuals and how they process, filter and understand messages. According to Calabrese, "the conceptual filters adopted by the individual players allow the structuring of chaotic environments and thus form the locus of the communication" [28]. However studying individuals' conceptual filters is difficult. In addition, the assumption of linear causality and its focus on physiological aspects of communication at the expense of the physical aspects also are pointed out as shortcomings of the model. The psychological models are more abstract and related to individuals. However, the Multiple Resource Model interprets communication in terms of available mental resources and multitasking and, therefore, may be particularly relevant to the context of this paper.

**Multiple Resource Model**

The Multiple Resource Model is the practical application of Wicken's Multiple Resource Theory which proposes that the "human operator does not have one single information processing source that can be tapped, but several different pools of resources that can be tapped simultaneously" [29]. The Multiple Resource theory originates from the concept of 'single channel bottleneck' and the `limited capacity central processor' (Moray) as well as the attention model proposed by Daniel Kahneman. The Multiple Resource Model?? is important because it addresses the issue of multitasking by making a structural distinction between auditory and visual processing in multitasking. According to the model, "Time-sharing between two tasks was more efficient if the two utilized separate structures than if they utilized common structures" [29].In the applied context, "the value of such models lies in their ability to predict operationally meaningful differences in performance in a multi-task setting, that results from changes (in the operator or in the task design) that can be easily coded by the analyst and the designer" [29].

Work inside a hospital is characterized by its chaotic nature; multitasking is very common. In psychological terms, multitasking is described as time sharing of internal resources between various modes of cognition such as visual, auditory and spatial. In addition to multitasking, healthcare professionals receive messages in multiple formats - visual, verbal, auditory or a combination of those. Wicken's Multiple Resource Model can be applied to evaluate healthcare workflows and to make efficient use of available resources.

**Systemic interaction model**

The Systemic interaction model establishes the communication process as a set of processes that must be analyzed as a whole before the communication can be understood[19].Unlike the mechanical approach, the systemic interaction approach does not assume that the entire communication process is equal to the sum of the individual acts. Rather, it is the communication process which determines the organizational behavior. The key to communication in the systemic interaction model is in the ordered rendering of structured behaviors discerned in terms of recurrence. The Systemic interaction model is not a search for cause effect analysis connecting the ingredients of the process of communication[19]. The systemic interaction model is closely related to systems theories. Like the systems theory, the systemic interaction model is highly relevant in today's hospitals because each unit in a hospital can be represented as a subsystem of the larger system (hospital).

**Schein's cultural model**

Although not strictly a communication model Edgar Henry Schein's cultural model is very much pertinent in complex sociotechnical systems such as hospitals. According to Schein, culture can be represented as a three layered model. Artifacts and behavior makes up the first level. The second level at which culture can be explained is found in espoused values. These are the values individuals claim they support. An important point is that espoused values can be in conflict with artifacts and behaviors. Basic assumptions, the final layer, culture, in Schein's model lie deeper than the other two levels and are the most difficult to discern. The assumptions are the set of fundamental beliefs which are engraved on to the collective inner psyche of a culturally distinct cluster of people[28]. The beliefs develop over long periods of time from the group's antecedents, what made them successful, or otherwise, and from the opinions and practices of their founders[30]. Attempts to improve safety culture within an organization as complex and diverse as modern hospitals need to afford prime importance to all the three layers.

**Westrum's Model**

Westrum's model is a broader approach towards representing communication. In this model, three communication styles (pathological, bureaucratic and generative) produce different organizational climates and handle safety information quite differently. In groups with a generative culture, hidden failures are actively sought and if possible, removed. However, generative culture can only be successful if the management not only encourages people at all levels to communicate but also urges all personnel to critically evaluate all levels of operation. These organizations that follow generative methods also develop effective ways of reporting problems and deal positively with errors; the system learns through its mistakes rather than punishing those that are involved in the error chain. The staff will generally increase their risk-taking behaviors in environments which follow extreme bureaucratic or pathological cultures. Undoubtedly hospitals need to model their communication strategies to move towards adopting generative culture. It is important for hospital administrators to understand that introduction of state-of-the-art communication devices cannot essentially change the culture within the organization. For effective communication to occur, introduction of newer and more efficient ICT enabled communication devices has to be coupled with changing the organizational culture[31].

**SMCR (Source, Message, Channel, Receiver) Model**

The SMCR model (S - source, M - message, C - channel, R- receiver), proposed by Berlo is an adaptation of Shannon Weaver's model. Berlo's model consists of the following 5 components:

1. The communication source—someone with a reason for engaging in communication; factors include communication skills, attitudes, knowledge level and position within a social cultural system;

2. The encoder who must have motor skills, such as vocal mechanisms that produce sound or muscle systems that produce gestures or writing;

3. The message—a systematic set of symbols;

4. The channel—a medium, a carrier of messages;

5. The decoder—the receiver's sensory skills that retranslates the message into a form he can use;

6. The communication receiver—the target of the communicator.

According to Speckhard, the SMCR model divides the communication process into source-receiver and receiver-source components; each having an encoder-decoder and a channel. The encoder-decoder unit performs the function of symbolizing or de-symbolizing the message flow in either direction. The source may be a receiver and vice versa. The direction of the message flow determines whether the encoding or decoding function is performed [32].

**Andersch, Staats and Bostom model**

The Andersch, Staats and Bostom model emphasizes environmental or contextual factors. This model, however, goes deeper than merely considering the effect of environment in communication processes. It gives emphasis to the transactional nature of communication in which the sender and the receiver are both interacting with the environment while the former is trying to convey a particular meaning and the latter trying to reconstruct and comprehend [30].

**Mechanical/ Mathematical Models**

These models were the earliest models of communication. According to mechanical models, communication is a linear sequential and simplistic process whereby the message is created, encoded, transmitted through a channel and decoded. Another important tenet of the mechanical model is that all language used in communication needs to be free of ambiguity. Apart from its linearity, the model has several shortcomings. It accords a quasi-causal relation to the sender and the receiver. Additionally, the model considers messages as objects with spatial and physical properties such as frequency, amplitude or duration. Any distortions are interpreted as "errors or noises to be eliminated as a result of the artificial belief that such errors or noises make it impossible to capture the nature of the message, which is considered neutral and previously defined" [28] .

**Shannon Weaver's Mathematical Model**

Shannon weaver's  model consists of five elements: an information source (creates a message), a transmitter, (encodes the message into signals) channel, receiver and destination[33].A sixth element, noise, was a later addition to the model. Whatever blocks or negatively affects the passage of the message through its channel is deemed to be a noise. The model looks at communication as a one-way process. This flaw was later corrected by adding a feedback process in the circular model of communication [31]. Although the Shannon Weaver's Mathematical model and its modifications are not relevant in modern day hospital communication studies it was included in the review because the model has been regarded as the precursor of all subsequent mathematical models and is a good starting point for anyone attempting to study any communication.

**Action, Interaction and Transactional Models**

The Action model suggests that communication is a simple process of injecting our messages into those we feel need to know the information. This model also has been referred to as linear communication model[4].

In the interaction model, feedback from the receiver is considered. The model acknowledges that communication is not strictly a one-way process with direct and linear effects. A transactional view of communication, like an interactional view, includes the important role of feedback. However, a transactional view goes further, in that it considers communication as a process with constant mutual influence of communication participants. The model's transactional view emphasizes the importance of context in the communication process. That is, not only do participants constantly influence each other, they are also influenced by the context in which they interact.

**Helical Model**

Dance's (1967) Helical model combines aspects of the early linear models of communication and the later circular models of communication to form a helix. Linear models omit the role of feedback in communication and circular models are flawed in that they suggest communication comes full circle to the same point from which it started. The helix implies that while communication is moving forward, it is also coming back upon itself and is being affected by its past behavior. For instance, what is communicated now will influence the structure and content of communication later on. The helix suggests that different aspects of the communication process change over time. The shape of the helix can differ according to the situation - with prior knowledge of a topic the helix widens quickly. With little or no prior knowledge, such as in new situations, the helix widens more slowly[31]. The helical model is likely more representative of hospital communication than the previous two linear and circular models. This is because the helical model overcomes certain disadvantages of the other two and suggests that the communication exchange is not a linear, one way process or a complete circle but a transactional, 2 way process.

**Schramm's Model**

The Schramm communication model is basically a modification of the Shannon-Weaver model and incorporates some of its components and other technical aspects of communication. Schramm was interested in the instructional role of communication.

Therefore, the primary concerns of his model are the meaning and the communication of meaningful symbols. The Shannon-Weaver model has little to say about meaning. Schramm maintained that encoding and decoding is not a linear sequential process, but performed simultaneously by the sender and the receiver. The Schramm model views communication as a two-way exchange of information.

Major strengths of the Schramm model are its inclusion of a 'field of reference' (psychological framework), feedback, context (a message may have different meanings, depending upon the specific context or setting) and culture (a message may have different meanings associated with it depending upon the culture or society). Communication systems, according to Schramm, operate within the confines of cultural rules and expectations to which we all have been educated) [34].

One drawback of the Schramm model is that it does not allow for complex, multiple communications between more than two persons. The inclusion of context, field of reference, culture and feedback makes the Schramm model more relevant to hospital communication than linear models.

**Conclusion and recommendations**

Our review has several limitations. First, although we have used comprehensive review methods across a large number of core biomedical databases and other sources the vastness of the knowledge areas to be mapped might have resulted in some pertinent results not being located by us. Second, several non-biomedical databases do not respond well to keyword based searches and might have contributed to the lack of complete retrieval.

To conclude, our review was able to identify a number of theories and models that may be applicable in furthering hospital communication research. However, it must be noted that most of the theories and models have not yet been applied in biomedical research to hospital communication and as such their applicability can only be suggested–a gap which future research may be able to address. Also we feel that there is significant scope for a new model representing the unique features and complexities of communication within hospitals.